\documentclass[%
 reprint,
 amsmath,amssymb,
prb,
]{revtex4-2}

\usepackage{graphicx}
\usepackage{dcolumn}
\usepackage{bm}
\usepackage{color}
\usepackage{multirow}

\DeclareUnicodeCharacter{2212}{-}

\begin{document}


\title{Computational insights into phase equilibria between wide-gap semiconductors and contact materials}

\author{Cheng-Wei Lee}
\affiliation{Colorado School of Mines, Golden, CO 80401,USA}
\affiliation{National Renewable Energy Laboratory, Golden, CO 80401,USA }
\author{Andriy Zakutayev}
\affiliation{National Renewable Energy Laboratory, Golden, CO 80401,USA }
\affiliation{Colorado School of Mines, Golden, CO 80401,USA}

\author{Vladan Stevanovi\'c}
\email{vstevano@mines.edu}
\affiliation{Colorado School of Mines, Golden, CO 80401,USA}
\affiliation{National Renewable Energy Laboratory, Golden, CO 80401,USA }


\begin{abstract}
Novel wide-band-gap (WBG) semiconductors are needed for next-generation power electronic but there is a gap between a promising material and a functional device. Finding stable (metal) contacts is one of the major challenges, which is currently dealt with mainly via trial and error. Herein, we computationally investigate the thermochemistry and phase co-existence at the junction between three wide gap semiconductors, $\beta$-Ga$_{2}$O$_{3}$, GeO$_2$, and GaN, and possible contact materials. The pool of possible contacts includes 47 elemental metals and a set of 4 common, $n$-type transparent conducting oxides (ZnO, TiO$_2$, SnO$_2$, and In$_2$O$_3$). We use first-principles thermodynamics to model the Gibbs free energies of chemical reactions as a function of the gas pressure (p$_{\mathrm{O}_2}$/p$_{\mathrm{N}_2}$) and equilibrium temperature. We deduce whether a semiconductor/contact interface will be stable at relevant conditions or a chemical reaction between them is to be expected, possibly influencing the long-term reliability and performance of devices. We generally find that most elemental metals tend to oxidize or nitridize and form various interface oxide/nitride layers. Exceptions include select late- and post-transition metals, and in case of GaN also the alkali metals, which are predicted to exhibit stable coexistence, although in many cases at relatively low gas partial pressures. Similar is true for the transparent conducting oxides, for which in most cases we predict a preference toward forming ternary oxides when in contact with $\beta$-Ga$_{2}$O$_{3}$ and GeO$_{2}$. The only exception is SnO$_2$, which we find to be able to form stable contacts with both oxides. Finally we show how the same approach can be used to predict gas partial pressure vs. temperature phase diagrams to help direct synthesis of ternary compounds. We believe these results provide a valuable guidance in selecting contact materials to wide-gap semiconductors and suitable growth conditions.
\end{abstract}

\maketitle


\section{Introduction}
It is challenging to find suitable contact materials for wide-band-gap (WBG) semiconductors, in particular for power electronic devices. Namely, while novel wide-gap (WG) and ultra-wide-gap  (UWG) semiconductors are critically needed as the basis for the next generation power conversion devices capable of meeting the demands of the expected broad future electrification and adoption of renewable energy technologies \cite{Huang_power_2017,Gorai_EES_2019,Garrity_PRXEnergy_2022}, having the promising active material alone is not sufficient. Each new WBG/UWBG semiconductor must be accompanied by suitable contact materials for the devices to operate as desired. 

For any given WBG/UWBG semiconductor, suitable contacts need to fulfill a number of criteria. The list usually starts with electronic properties including electric conductivity of a certain magnitude, favorable Schottky barrier and/or or band alignment with the active material, lack of problematic interface states, etc. \cite{Sze_book, Tsao_AEM_2018}. However, many of these relevant quantities depend on the details of interface chemistry and the formation of secondary phases (or not) between active materials and contacts. Regardless whether the interface chemistry is helpful, as may happen in some instances, or harmful for the device performance, the knowledge of what exact phases are present at the interface is vital for all aspects of device engineering. This is particularly true for high-power and/or high-temperature electronic devices as the Joule heating due to large current densities in combination with frequent switching will likely lead to the equilibration of the system over long periods of time.

\begin{figure}[!t]
\centering
\includegraphics[width=\linewidth]{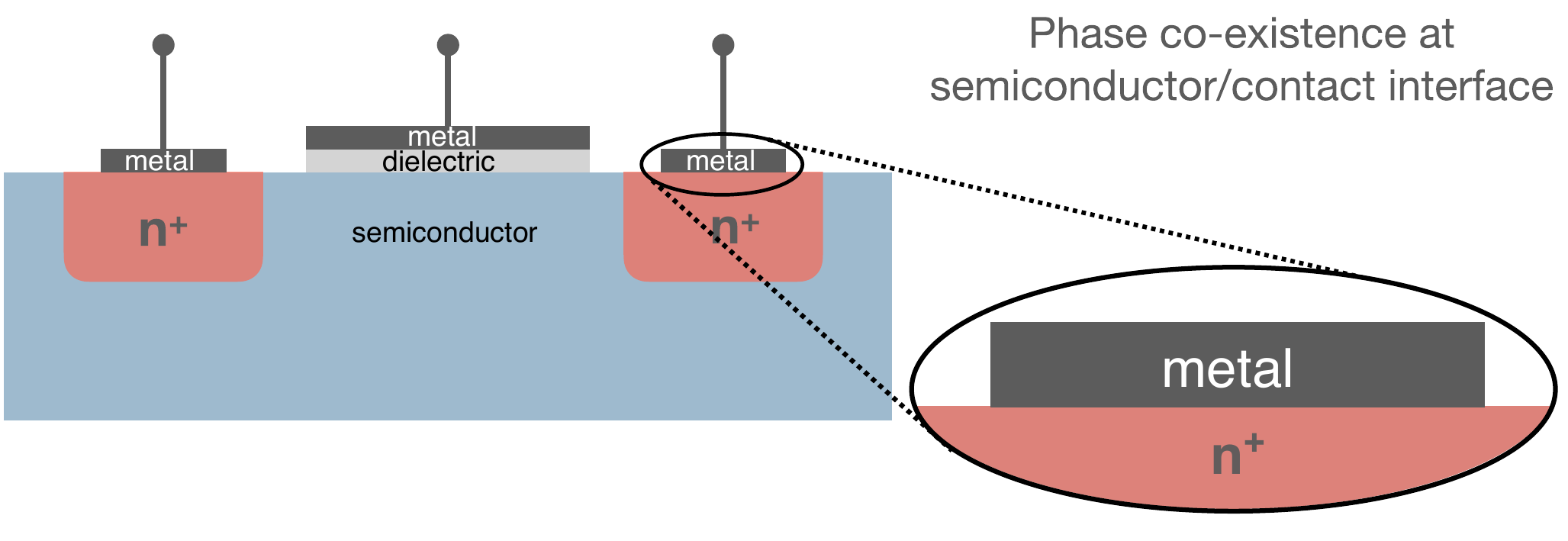}
\caption{\label{fig:intro} 
This paper discusses the chemistry and phase co-existence at metal/semiconductor interfaces and the implications related to finding suitable and chemically stable contacts to wide-gap semiconductors.}
\end{figure}

Interface chemistry or a stable coexistence between two materials in contact with one another can be evaluated from: (a) the knowledge of chemical reactions that could potentially happen, and (b) the Gibbs free energies of formation of all reactants and products of those chemical reactions. In case of oxidation-reduction reactions this could be done using the Ellingham diagrams for example \cite{gaskell2017introduction}. However, experimental thermochemical data, including the Gibbs free energies of formation ($\Delta G_f$) as well as the enthalpies of formation ($\Delta H_f$) for ternary and other multinary compounds are not as available as for the binary ones \cite{Stevanovic_PRB_2012,Kirklin_npj_CM_2015}. Hence, the chemistry that is likely to occur at the interface between WBG/UWBG materials, that are often binary compounds, and their contacts, often elemental metals or compounds themselves, cannot be generally predicted solely from the experimental data. 

In order to study the possible chemical reactions at the junction between WBG/UWBG semiconductors and their contacts in this paper we utilize computational resources and methods that allow interface chemistry and phase coexistence (i.e., the phase equilibria) to be evaluated more broadly and for larger range of compounds. First, we use the calculated enthalpies of formation \cite{Stevanovic_PRB_2012,Ong_CM_2008} stored in various computational materials databases, in combination with the modeled phonon contributions to the Gibbs free energies following the work by Bartel \emph{et al.} \cite{Bartel_NC_2018}. In this way predictions of the temperature-dependent, compound $\Delta G_f$ values for virtually any stoichiometric and ordered compound can be made and used to compute the reaction free energies. Second, the data stored in computational materials databases cover not only experimentally realized ternary and other multinary compounds but also the hypothetical ones that could potentially form (see for example Refs.~\cite{Sun_Nature_Materials_2019,pandey_pattern_2021}), providing the unprecedented list of possible solid phases, and hence possible reaction products. 

With the use of these databases, primarily the NREL computational materials database (NRELMatDB) \cite{nrelmatdb} and the Materials Project \cite{Jain_APL_Materials_2013}, we assess the interface chemistry and phase co-existence of 47 elemental metals and 4 commonly used transparent conducting oxides when in contact with $\beta$-Ga$_{2}$O$_{3}$, rutile GeO$_{2}$, and GaN. This set of materials covers novel WBG semiconductors ($\beta$-Ga$_{2}$O$_{3}$), recently proposed ones (rutile GeO$_{2}$ \cite{Chae_APL_2019}) as well as a well-established wide-gap semiconductor (GaN). 
In short, we find that most elemental metals tend to form various interface oxide/nitride layers. Exceptions include select late- and post-transition metals, and in case of GaN also the alkali metals, which are predicted to exhibit stable coexistence. Contrary to Ga$_2$O$_3$ and GeO$_2$ for which stable co-existence with most of those elemental metals occurs at relatively low gas partial pressures, junctions between GaN and metals are predicted to survive to high N$_2$ pressures owing to the strong triple bond in the N$_2$ molecule. 
Similar is true for the TCOs, for which in most cases we predict a preference toward forming ternary compounds with Ga$_2$O$_3$ and GeO$_2$. The only exception is SnO$_2$, which can co-exist with both Ga$_2$O$_3$ and GeO$_2$ and form a stable contact. In what follows we describe the methods and results in greater detail, and discuss the guidelines distilled from theory in choosing contact materials and appropriate growth conditions.
%
\section{Methods}
%
\subsection{Predicting thermochemistry and phase equilibria}\label{sec:stability}
We use the grand canonical ensemble formalism to evaluate the driving forces for interface chemistry and predict phase equilibria at the semiconductor/contact interface. This is necessary when dealing with elemental constituents such as O and N that are normally in the gaseous state.  Briefly, for a system in equilibrium with its environment and elemental chemical reservoirs, its state can be described in terms of chemical potentials of constituent elements, temperature and pressure. 
Any choice of pressure and temperature will determine the values of all elemental chemical potentials and the Gibbs free energies of all compounds within a given chemical system. The state of the system will be the one that minimizes the total Gibbs free energy including all relevant phases. This minimal condition can be elegantly formulated using the following system of inequalities \cite{Stevanovic_PRB_2012}:
\begin{gather}\label{eq:conv_hull}
\sum_{j=1}^{C} \, n_{ij} \, \Delta\mu_j (T,p) \,\, \leq \,\, \Delta G_f^{(i)} (T,p),\\ \nonumber
i=1,\dots, P. \nonumber
\end{gather}
Where $j$ counts different components (or elements, totaling to $C$), $i$ counts all possible phases the could form (totaling to $P$), $n_{ij}$ are the number of atoms of the component $j$ in the chemical formula of the phase $i$, $\Delta\mu_j (T,p)$ are the elemental chemical potentials expressed relative to the chemical potential of the same component in its standard state ($\mu_i$) at the respective temperature ($T$) and pressure ($p$), and $\Delta G_f^{(i)} (T,p)$ are the Gibbs free energies of formation of each phase expressed per one formula unit. It is important to note that all pure elemental phases are also included in these inequalities. Namely, for each element there will be only one $n=1$ with all others zero, and by definition $\Delta G_f=0$ for each pure elemental phase, implying $\Delta\mu_j \leq 0$. 

Also, while strict inequalities can be all fulfilled simultaneously, not all \emph{equalities} can exist simultaneously. The \emph{equalities} that are fulfilled for a given set of $\Delta\mu_j \leq 0$ values determine which phases are stable at those conditions. If only one equality is satisfied for a given set of $\Delta\mu_j$ values, this implies the thermodynamic stability of the corresponding compound, and the instability of all others for which $\sum n_{ij} \, \Delta\mu_j < \Delta G_f^{(i)}$. Two or more equalities existing simultaneously imply coexistence of the corresponding phases at those conditions.

The set of inequalities from eq.~\eqref{eq:conv_hull} defines a convex hyper-polygon in the chemical potential (hyper)space whose faces represent the single phase regions, and vertices and edges represent the phase coexistence.  To solve the inequalities \eqref{eq:conv_hull} we use the Double Description Method by Motzkin et al. \cite{Motzkin_ddlib} implemented as the C library cddlib. To include as many states as possible we utilize existing computational databases, specifically NRELMatDB \cite{Stevanovic_PRB_2012,nrelmatdb} and Materials Project \cite{Jain_APLMater_2013}, which account for most of the experimentally realized compounds, along with hypothetical ones. Additionally, in all our considerations we assume a low-pressure ($p\approx 0$) conditions for all solid phases. Pressure dependencies are included in the chemical potentials of the gaseous species, which depend more strongly on the corresponding partial pressures. Considered gas partial pressures are all equal or below 1 atm., which remains practically zero for solid phases. The dependencies of the elemental chemical potentials of gaseous species on the respective partial pressures are approximated using the ideal gas law. 

\begin{figure}[!t]
\centering
\includegraphics[width=\linewidth]{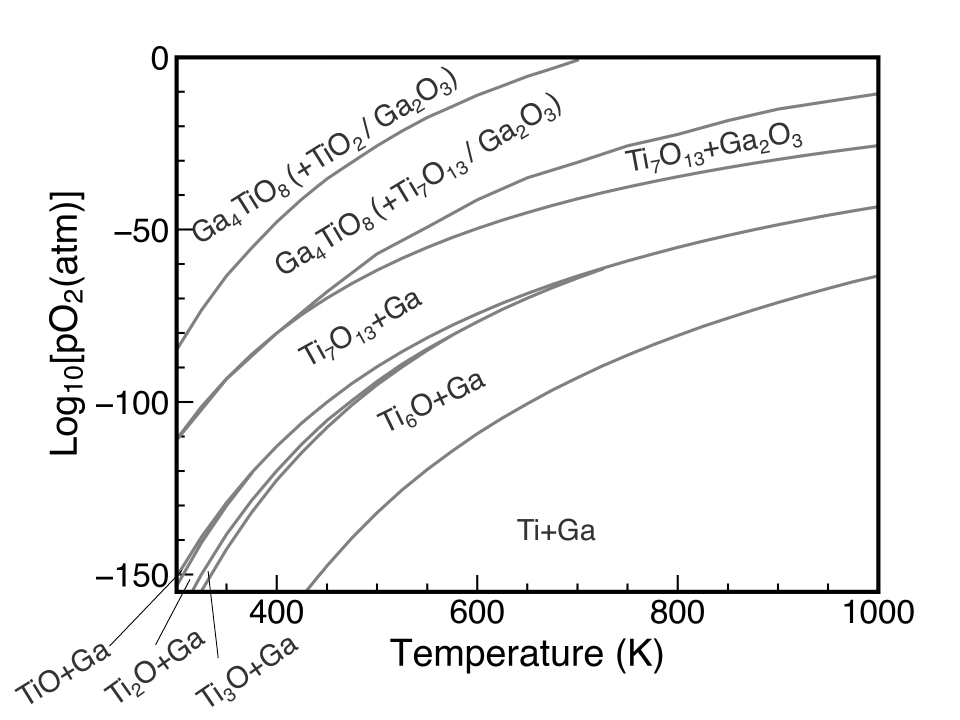}
\caption{\label{fig:Ga2O3_Ti} 
Predicted pO$_{2}$-T phase diagram for the Ga-Ti-O chemical system. Each region is labeled by the compounds that are predicti to form at those conditions. Compounds in brackets are those predicted to form when the Ga:Ti ratio deviates from 4:1; depending whether there is an excess of Ti or Ga different compound will form (separated by ``$/$'' sign). Large range of pO$_{2}$ values is shown for completeness. 
}
\end{figure}

To illustrate our approach let's consider Ga-Ti-O chemical system that is of relevance for making ohmic contacts to $\beta$-Ga$_{2}$O$_{3}$ \cite{Callahan_JVSTA_2023}. When hcp Ti is deposited on $\beta$-Ga$_{2}$O$_{3}$, the stability of  Ti/Ga$_{2}$O$_{3}$ interface at given conditions depends on whether a combination Ti + Ga$_{2}$O$_{3}$ has the lowest Gibbs free energy among all the possible compounds that are made of Ga, Ti, and O. To answer this question we consider all possible compounds (phases) within this chemical space including Ga$_{2}$O$_{3}$ and TiO$_{2}$, but also the ternary compound Ga$_{4}$TiO$_{8}$ and the Ti-O Magnelli phases  Ti$_{9}$O$_{17}$, Ti$_{8}$O$_{15}$, Ti$_{7}$O$_{13}$, Ti$_{6}$O$_{11}$, Ti$_{5}$O$_{9}$, Ti$_{4}$O$_{7}$, Ti$_{3}$O$_{5}$, Ti$_{2}$O$_{3}$,Ti$_{4}$O$_{5}$, TiO, Ti$_{2}$O, Ti$_{3}$O, and Ti$_{6}$O. Results are summarized in the form of the pO$_{2}$-temperature phase diagram shown in Fig.~\ref{fig:Ga2O3_Ti}.

Each region is labeled by a single phase or two phases that can coexist. Labels in parentheses refer to two possible options for the phase coexistence. For example, in the top left part of the diagram the Ga$_4$TiO$_8$ ternary compound could stably coexist at those conditions either with TiO$_2$ or Ga$_2$O$_3$ (not both simultaneously) depending on the $\Delta \mu_{Ti}$ and $\Delta \mu_{Ga}$, or, in other words, the excess of one or the other relative to the perfect $\mathrm{Ga}: \mathrm{Ti}=4:1$ ratio. In the case of a perfect $4:1$ only the Ga$_4$TiO$_8$ ternary would form. Lines in Fig.~\ref{fig:Ga2O3_Ti} represent the three phase coexistence. 

It is evident from this plot that $\beta$-Ga$_{2}$O$_{3}$ is never in equilibrium with elemental Ti, always with one of its oxides or the ternary compound, implying that the Ohmic contact typical for $\beta$-Ga$_{2}$O$_{3}$ either has a secondary oxide layer that forms spontaneously or is in the metastable state that is likely going to evolve with time. Experimental evidence supporting this prediction can be found in the our recent work\cite{Callahan_JVSTA_2023}, where it was shown that Ti oxidizes almost fully when in contact with $\beta$-Ga$_{2}$O$_{3}$ leading to the degradation of the device performance over time when exposed to elevated temperatures. What this type of analysis requires are the $\Delta G_f$ values and their temperature dependence for all phases including the elemental chemical potentials. How are $\Delta G_f$ obtained in our approach is discussed next.

\subsection{$\Delta G_f$ of inorganic compounds}
The Gibbs free energy (molar) of a given compound is a function of temperature, and pressure. Most computational materials databases however, only report low temperature and $p=0$ enthalpies of formation ($\Delta H_f$), which can serve as the approximation for the low temperature and low pressure $\Delta G_f$ values. In addition, common approximations to density functional theory (DFT), like the GGA for example, typically result in large differences between the computed $\Delta H_f$ values of inorganic compounds and the measured ones (on average by $\sim$250 meV/atom \cite{Stevanovic_PRB_2012}). To address this limitation of DFT, we primarily use the $\Delta H_f$ values from NRELMatDB  calculated using the fitted elemental-phase reference energies (FERE) approach of Stevanovi\'c \emph{et al.} \cite{Stevanovic_PRB_2012}, which allows improved predictions of the compound formation enthalpies ($\approx$ 50 meV/atom). Similarly, the data from Materials Project employs an alternative way of correcting for the DFT deficiencies, which combines DFT and DFT+U \cite{Jain_PRB_2011}.

For the temperature dependence of the Gibbs free energies, we utilize the model description of the phonon contributions to the $\Delta G_f$ developed by Bartel \emph{et al.} that was shown to result in the temperature dependent $\Delta G_f$ values with the accuracy of $\approx$50 meV/atom that is largely inherited from the $\Delta H_f$ values that are used as the starting point \cite{Bartel_NC_2018}. The model equation from Bartel et al., is:
\begin{equation}
\Delta G_{f}^{(i)}(T) = \Delta H_{f}^{(i)} + \delta G^{(i)}(T)-\sum_{j}^{N} n_{i,j}\mu_{j,exp.}(T),
\end{equation}
where $\Delta H_{f}^{(i)}$ represents the enthalpy of formation of the phase $i$ at standard conditions (from FERE), $\delta G^{(i)}(T)$ is the term that introduces the temperature dependence modeled in the work of Bartel \cite{Bartel_NC_2018}, as before $n_{i,j}$ are the stoichiometric weights of elements, and $\mu_{i,exp.}(T)$ are the temperature dependent experimental chemical potentials of elemental constituents in their reference phases that are taken from the FactSage package \cite{Bale_factsage_2016}. When calculating $\Delta G_{f}^{(i)}(T)$ the most stable (lowest Gibbs free energy) phases are used for elements if phase transitions occur with increasing temperature. The $\delta G^{(i)}(T)$ (in eV/atom) is defined by,
\begin{gather} 
\delta G^{(i)}(T)  = \left(-2.48 \,\, 10^{-4} \,\, ln(V) \, - \, 8.94 \,\, 10^{-5} \, \frac{m}{V} \right) \, T \, + \, \\ \nonumber
                                         +\,\, 0.181 \,\, ln(T) \, - \, 0.882, \\ \nonumber
\end{gather}
where $V$ is the volume of the compound  in {\AA}/atom and $m$ is the reduced atomic mass (in amu).

For the two gaseous molecules, O$_{2}$ and N$_{2}$, considered in the analysis, we modeled the dependence of their relative chemical potentials ($\Delta\mu_i$), to the temperature T, and gas partial pressure using the ideal gas law, $\Delta \mu_{O}= 1/2 \, k\mathrm{_{B}} \, T \, ln(p\mathrm{_{O_{2}}}$) and $\Delta \mu_{N}= 1/2 \, k\mathrm{_{B}} \, T \, ln(p\mathrm{_{N_{2}}}$) for O$_{2}$ and N$_{2}$ referenced to the experimental value value $\mu_{i,exp.}(T)$ at standard pressure (1 atm). As a result, we can construct the p$\mathrm{_{O_{2}}}$-T and p$\mathrm{_{N_{2}}}$-T phase diagrams as the one in Fig.~\ref{fig:Ga2O3_Ti}. 

Lastly, we also estimate the uncertainty of the p$\mathrm{_{O_{2}}}$-T phase diagram predictions. For the same amount of uncertainty in chemical potential of O or N, the ideal gas model indicates that the uncertainty in partial pressure strongly depends on temperature. For instance, an overestimate of oxygen chemical potential by 50 meV will lead to about 50 times larger oxygen partial pressure at 300 K but it only gives rise to about 4 times larger oxygen partial pressure at 900 K. Therefore, even though the uncertainty in chemical potential of gaseous elements is system-dependent, we can generally expect lower uncertainty in predicted partial pressure at higher temperature. While these uncertainties may seem large, which is a consequence of the exponential dependency on chemical potential, the order of magnitude of the gas partial pressures is robustly predicted at typical synthesis temperatures. The results presented here hence, need to be interpreted as predictive in the order magnitude of partial pressure at any given temperature.

\section{Results and discussion}
%
\subsection{Stability of elemental metals in contact with wide gap semiconductors.}
%
\begin{figure}[!htb]
\centering
\includegraphics[width=0.99\linewidth]{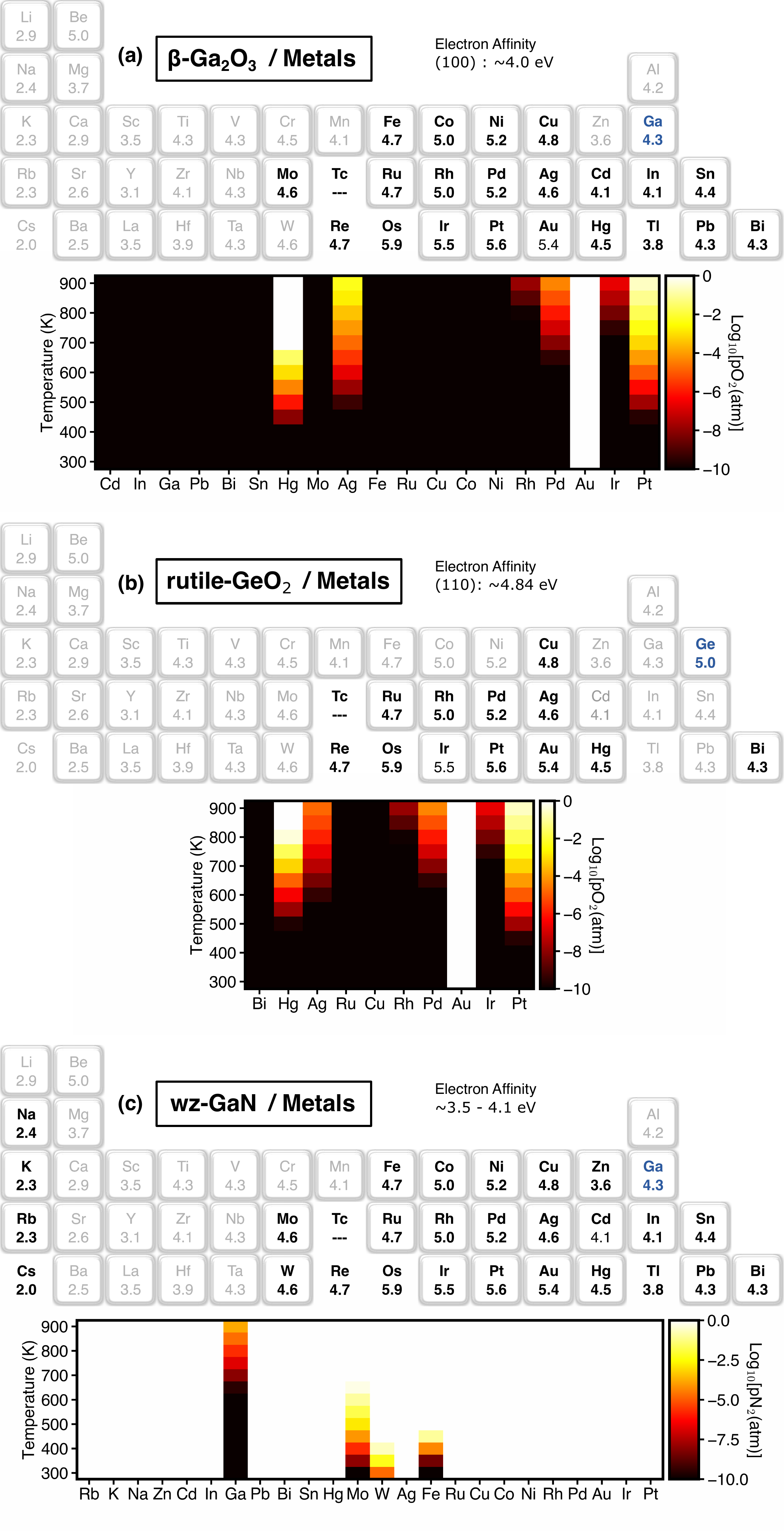}
\caption{\label{fig:Ga2O3_PT} 
Predicted stability of elemental metals in contact with: (a) $\beta$-Ga$_{2}$O$_{3}$ (c) rutile GeO$_{2}$, and (c) wurtzite GaN at 300 K and any gas partial pressure. Their predicted maximally allowed gas partial pressure (pO$_{2}$ or pN$_{2}$) for stable metal contacts are also shown. Numbers below the element symbols are metal work functions (see details in main text). Stable and unstable interfaces are in black and gray, respectively. Elements without boxes are predicted using Materials Project data \cite{Ong_CM_2008,Jain_APLMater_2013}. Experimental value for electron affinity of the Ga$_2$O$_3$ (100) surface is from Ref.~\cite{Mohamed_APL_2012}. 
}
\end{figure}
Using the approach described above we created the gas partial pressure vs. temperature phase diagrams as the one shown in Fig.~\ref{fig:Ga2O3_Ti} for a range of elemental metals and three wide gap semiconductors Ga$_2$O$_3$, GeO$_2$ and GaN.
The results are summarized in Fig.~\ref{fig:Ga2O3_PT}. Sections of the periodic table show all the elemental metals we investigated and the three panels correspond to the three wide gap systems. Stability information is presented in the following way. Chemical symbols in black correspond to the metals that are predicted to stably co-exist with each with gap system at 300 K and some gas partial pressure. The effect of temperature increase on stability assessment is that maximal partial pressure at which a given metal starts to oxidize or nitridize will shift to higher values. This effect can be observed in the bottom sub-panels of each panel, where the maximal gas pressures for the stable co-existence are shown as a function of temperature for the subset of metals that are declared stable at 300 K. Black color denotes maximal gas pressure values equal or below the $10^{-10}$ atm. 
In contrast, metals that are predicted to form compounds (binary, ternary, etc.) at 300 K and any gas pressure when in contact with our wide gap systems are shown in grey and declared unstable. This classification is robust with respect to changing conditions as those metals that are predicted unstable at all gas pressure at 300 K are not going to become stable at temperatures above 300 K and any physically reasonable gas pressures. 

It becomes clear that for Ga$_2$O$_3$ and GeO$_2$ every metal would oxidize above a certain gas pressure, and the pressures before this happens are below $10^{-10}$ atm for most of elemental metals. Only for a subset of metals, the noble ones mainly (plus Hg), we find oxidation to occur at $p\gtrsim10^{-10}$. The situation is markedly different in case of GaN. First, a larger range of elemental metals are predicted to stably co-exist with GaN at 300 K including nearly the entire right half of the periodic table and alkali metals (Na, K, Rb, Cs). This result is a consequence of the much lower number of binary and ternary nitride compounds in comparison to the oxides, which results from the much lower and even positive enthalpies of formation. The work of Sun et al. \cite{Sun_Nature_Materials_2019} reveals that no new ternary nitrides composed of Ga and other elemental metals are thermochemically stable, which is also true for the corresponding binary nitrides most of which have positive enthalpies of formation (see in NRELMatDB and Materials Project). Second, most of the metals that are predicted to stably co-exist with GaN can survive to relatively high N$_2$ pressures. Only Mo, W, and Fe are predicted to form nitride compounds below $\sim$500 K.

\subsection{Metal contact selection based on stability and metal workfunction}
Now we turn to discussing the type of contacts (Schottky or Ohmic) metals that are predicted to stably co-exist at some gas pressures and temperatures would form with the three WBG semiconductors. To find an elemental metal that would stably coexist and form an Ohmic contact with the $n$-type $\beta$-Ga$_{2}$O$_{3}$, a metal with work function (WF) around 4.0 eV is needed. This is based on the qualitative Schottky-Mott rule \cite{Sze_book} and the value for the electron affinity of $\beta$-Ga$_{2}$O$_{3}$ that is $\approx$4.0 eV for the (100) surface \cite{Mohamed_APL_2012}. Given the largely qualitative nature of the Schottky-Mott rule we set the target metal WF  to be  lower than 4.5 eV, which singles out Ga, Cd, In, Tl, Pb, and Bi as the potential stable metal Ohmic contacts for $\beta$-Ga$_{2}$O$_{3}$. However, there are two limitations regarding these metals. First, they all have low melting points ($<350^{\circ}$C) and such property prevents them from high temperature applications ($>400^{\circ}$C). Furthermore, they generally require low oxygen partial pressures ($<10^{-10}$ atm.) to maintain a stable metal/$\beta$-Ga$_{2}$O$_{3}$ interface. Hence, our stability analysis suggests that stable ohmic contacts at typical operating and/or synthesis conditions are unlikely to be found among elemental metals.

Qualitatively similar results are obtained in case of rutile GeO$_{2}$, which was proposed recently to be a promising wide gap semiconductor material \cite{Chae_APL_2019}. Metals that are predicted to form stable interface with rutile  GeO$_{2}$ generally have high work functions. Among them, only a few noble metals (Au, Hg, Pt, Ag, Pd, Ir, and Rh)are predicted to form stable  GeO$_{2}$/metal interfaces under practical operation condition (pO$_{2}$ $>$ 10$^{-10}$ atm). Considering the predicted electron affinity of $\sim$4.84 eV calculated by HSE06+G$_{0}$W$_{0}$, Ag and Ru, with work functions of 4.6 and 5.0 eV respectively, are predicted to be stable metal Ohmic contacts for the $n$-type doped rutile GeO$_{2}$, albeit at very low pO$_{2}$. Rutile GeO$_{2}$ is also predicted to exhibit ambipolar doping behavior \cite{Chae_APL_2019}, i.e. rutile GeO$_{2}$ can be not only $n$- but also $p$-type conductor. In the case of $p$-type doping and with predicted ionization potential of the rutile GeO$_2$ of 9.5 eV for the (110) surface, there are practically no metals that can form metal Ohmic contact according to the Schottky-Mott rule. 

Lastly, we discuss wurtzite GaN, which has different thermochemical properties from two WBG oxides. Fig.~\ref{fig:Ga2O3_PT} shows that alkali metals, except for Li, can form stable interfaces with GaN.  This observation is unintuitive at first glance given that alkali metals are known to be reactive with oxygen molecules. Further investigations into the phase equilibria of Ga-N-alkali systems show that only Li can form stable ternary compound (Li$_{3}$GaN$_{2}$) that prevents coexistence between Li and GaN. Heavier alkali metals (M=K, Na, K, Rb, Cs) can form MN$_{3}$ compounds but the formation energies are not low enough to prevent coexistence with GaN. 

\begin{figure}[ht!]
\centering
\includegraphics[width=\linewidth]{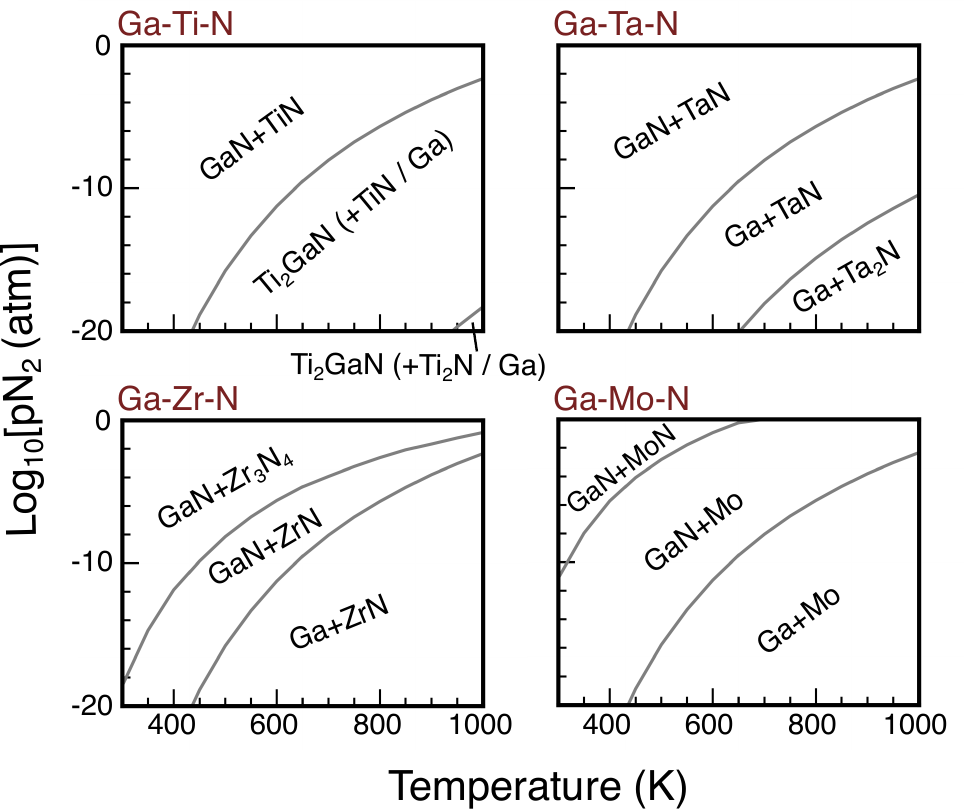}
\caption{\label{fig:PT_nitrides} 
Selected gas partial pressure $\emph{vs.}$ temperature phase diagrams that are relevant for GaN. Compounds within brackets are competing binary phases for the ternary compounds of interest and "/" suggests two options based on synthesis conditions.
}
\end{figure}
\begin{figure*}[!bht]
\centering
\includegraphics[width=0.9\linewidth]{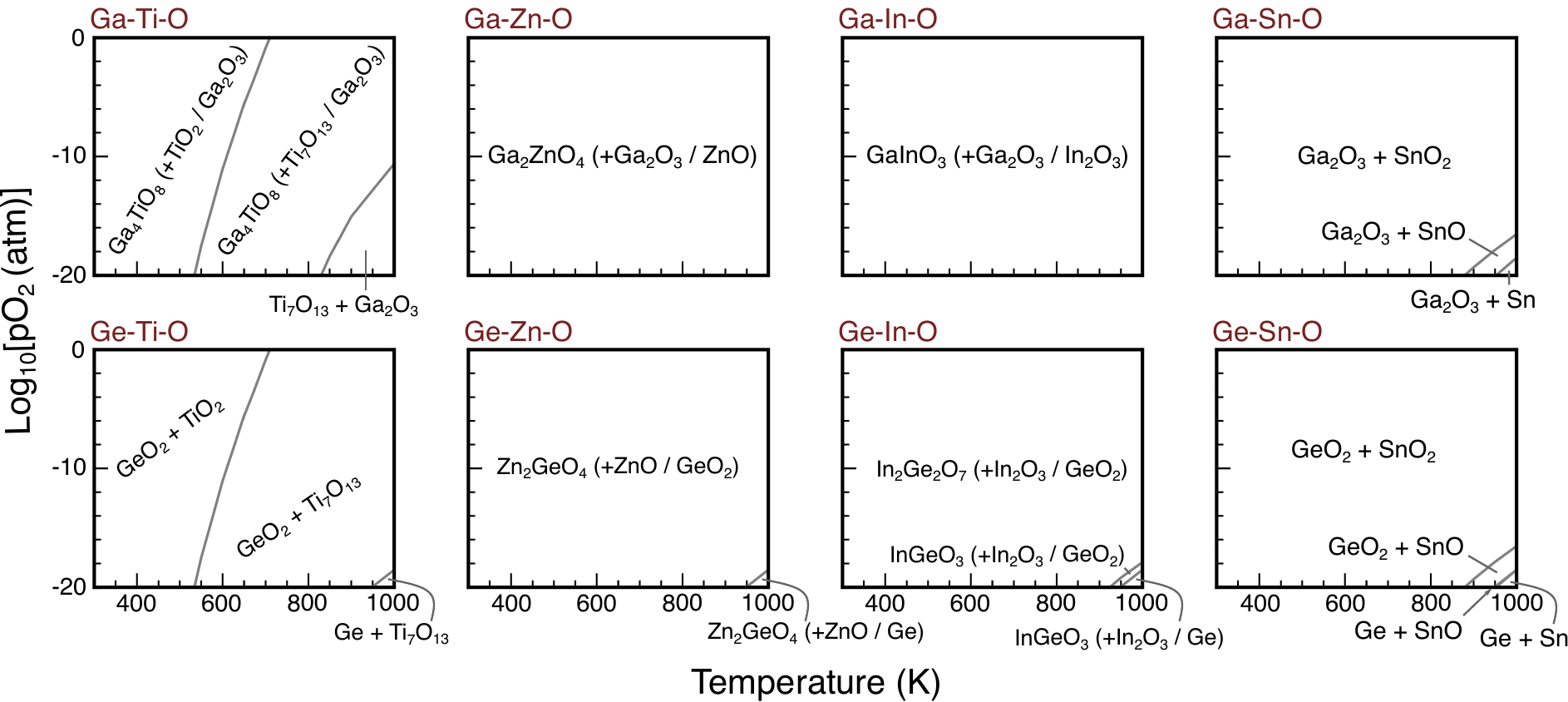}
\caption{\label{fig:PT_oxides}  
Selected gas partial pressure $\emph{vs.}$ temperature phase diagrams that are relevant for Ga$_{2}$O$_{3}$ and GeO$_{2}$. Compounds within brackets are competing binary phases for the ternary compounds of interest and "/" suggests two options based on synthesis conditions.
}
\end{figure*}

Considering the extended use of Ti as metal contact for GaN in the literature\cite{Greco_ASS_2016}, we performed a detailed analysis on the predicted results of GaN/Ti interfaces. Fig.~\ref{fig:PT_nitrides} shows that wz-GaN/Ti interfaces are not stable and further investigations into phase equilibria show that Ti will form TiN, or Ti$_{2}$GaN both of which have low enough formation energy to prevent coexistence between Ti and GaN. However, we note that these are conductive metal nitrides and the conductivity at the interface is likely still high irrespective of the formation of various Ti-N phases. Ta and Mo  are another two common metal contacts for GaN\cite{Greco_ASS_2016} and Fig.~\ref{fig:PT_nitrides} shows that wz-GaN/Ta interface is not stable, while the wz-GaN/Mo is predicted to be stable in the middle part of the corresponding phase diagram in Fig.~\ref{fig:PT_nitrides}. Similar to TiN, TaN and MoN are also electrically conductive.  Lastly, we look at stability of ZrN/GaN interface.  ZrN is a conductive nitride commonly used as metal contact and diffusion barrier for GaN\cite{Wolter_ECSSL_1999,Voss_APL_2007}.  We notice that ZrN/GaN interface can be stable at certain synthesis condition but at lower temperature and/or higher nitrogen partial pressures Zr$_3$N$_4$ will form, which is instead a semiconductor. This suggests that control of nitrogen partial pressure is needed to maintain long-term stability of ZrN/GaN interfaces. 

\subsection{Transparent conducting oxides as Ohmic contacts to Ga$_2$O$_3$ and GeO$_2$}

Given the apparent absence of elemental metals that would stably co-exist with $\beta$-Ga$_2$O$_3$ and form Ohmic contacts, we examine the possibility of using transparent conducting oxides in their place. The considered TCOs, including TiO$_2$, ZnO, In$_2$O$_3$ and SnO$_2$, are generally dopable to high electron concentrations and, since they are oxides themselves, it would be expected for them to form stable co-existence with other WBG oxides. This expectation is under examination here. If stable, TCOs may provide a solution to the Ohmic contact problem because when used as a buffer layer between the WBG semiconductor and an elemental metal, the highly doped TCO may help create an effective Ohmic contact due to the high charge carrier concentration and/or suitable band alignment. 

Our results illustrated in Fig.~\ref{fig:PT_oxides} show that neither ZnO nor In$_2$O$_3$ can co-exist with $\beta$-Ga$_{2}$O$_{3}$ but would rather form ternary oxides. Formation of Ga$_2$ZnO$_4$ spinel and GaInO$_3$ compound, respectively, is predicted to occur when in contact with $\beta$-Ga$_{2}$O$_{3}$. However, it is important to note that Ga$_{2}$ZnO$_{4}$ itself is a $n$-type material \cite{Chi_MTP_2021} and that formation of Ga$_{2}$ZnO$_{4}$ might not  be necessarily bad for the device performance, although the interface characteristics will depend on the chemistry. The case of Ga-Ti-O chemical system that was already discussed in the context of Fig.~\ref{fig:Ga2O3_Ti} deserves a little more attention. Namely, while the conclusion that $\beta$-Ga$_{2}$O$_{3}$ dos not stably co-exist with elemental Ti is robust, the conclusion suggested by the Fig.~\ref{fig:PT_oxides} that there is no coexistence with TiO$_2$ either may be questionable. This is because the reaction free energy between TiO$_2$ and $\beta$-Ga$_{2}$O$_{3}$ to form Ga$_4$TiO$_8$ is predicted to be only $\sim -11$ meV/atom at 300 K, which is within the error bar of our method. Similarly, the formation of GaInO$_{3}$ is also within $\sim$ 10 meV/atom. Consistent results can be obtained from Materials Project and OQMD databases. Hence, the Ga$_4$TiO$_8$ and GaInO$_3$ are only marginally stable according to our calculations and as a result, the stable coexistence with $\beta$-Ga$_{2}$O$_{3}$ may be a function of specific conditions at the interface that go beyond bulk thermodynamics (e.g. strain, local fluctuations of thermodynamic parameters etc.). 

In conclusion, the only TCO out of the 4 that we find to robustly co-exist with $\beta$-Ga$_{2}$O$_{3}$ without potentially forming ternary oxides is SnO$_{2}$.  For the rutile GeO$_2$ we find a stable coexistence with both TiO$_2$ and SnO$_2$ while the formation of ternaries Zn$_2$GeO$_4$ and In$_2$Ge$_2$O$_7$ is predicted to occur for ZnO and In$_2$O$_3$. The formation of the ternary compounds is in this case outside of the error bars of our methodology and these predictions are considered robust. 
The difference in electron affinities between SnO$_2$ ($\sim$5.3 eV\cite{Stevanovic_PCCP_2014}), Ga$_2$O$_3$ ($\sim$4.0 eV\cite{Mohamed_APL_2012}), and Ge$O_2$ ($\sim$4.8 eV) are also relatively large, which suggests that engineering of the Ohmic type contact by adjusting electron concentrations (doping) and the corresponding Fermi levels in both Ga$_2$O$_3$/GeO$_2$ and SnO$_2$ is likely required. This will, of course, depend on the details of the atomic structures at the interface.

\subsection{Phase diagrams for synthesis of ternary compounds}\label{sec:synthesis}
The surge in computational materials discovery has led to a growing candidate list of ternary and multinary compounds, which are generally more challenging to grow due to larger parameter space than binary compounds. The gas partial pressure versus temperature phase diagrams generated computationally are also useful tools to reduce the synthesis parameter space. Taking the ternary compounds in Fig.\ \ref{fig:PT_nitrides} and \ref{fig:PT_oxides} as examples, we can deduce that it will be simpler to grow single-phase Ga$_2$ZnO$_4$, GaInO$_3$, and Zn$_2$GeO$_4$ than Ti$_2$GaN, Ga$_4$TiO$_8$, and In$_2$Ge$_2$O$_7$. This is based on the complexity of phase diagram at the shown range of partial pressure and temperature. For the former three ternary compounds, the synthesis is expected to focus on tuning the cation ratio avoid secondary phases. However, for the latter three compounds synthesis requires further control of gas pressure and temperature. Specifically, lower temperature  is needed to grow Ga$_4$TiO$_8$ while higher temperature is needed for Ti$_2$GaN. To prevent growing oxygen-deficient phase (InGeO$_3$) for In$_2$Ge$_2$O$_7$, high-temperature and low oxygen partial pressure should be avoided.

\subsection{Qualitative correlation between the metal-oxide formation enthalpies and the metal work functions}\label{sec:eform}
%
Based on thermochemical stability predictions for metal/oxide interfaces studied in this work, we noticed the following trend. Metals with smaller work functions, i.e. Fermi level closer to vacuum, are generally less stable when in contact with $\beta$-Ga$_2$O$_3$ or GeO$_2$. In other words their oxides are more likely to form than oxides of the metals with higher work functions. Furthermore, using $\beta$-Ga$_{2}$O$_{3}$/metal coexistence as an example, we found that metals which are more likely to form oxides when in contact with $\beta$-Ga$_{2}$O$_{3}$ generally have their work functions lower than the work function of elemental Ga ($\sim$4.3 eV). These two observations indicate that metals with lower work functions in general have more negative oxide formation energy and thus have stronger tendency to form oxides when two metals compete for forming metal-oxygen bonds. 

\begin{figure}[!thb]
\centering
\includegraphics[width=0.9\linewidth]{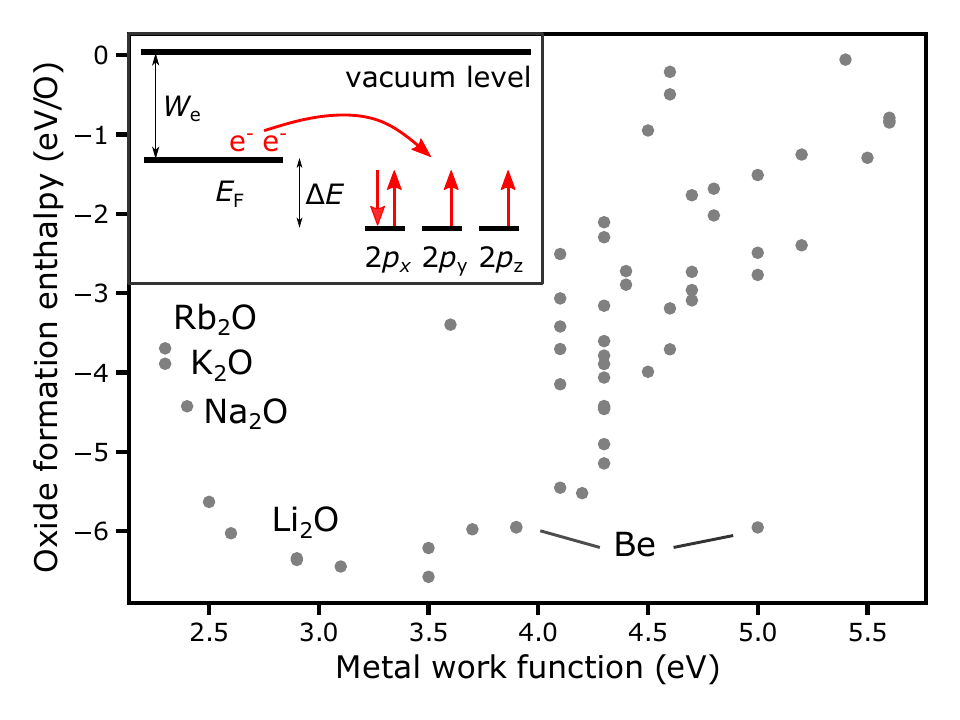}
\caption{\label{fig:Eform_EWF} 
Correlation between formation enthalpy of metal oxides per oxygen atom and metal work function of the constituent metals. 
The inset illustrates the underlying chemical intuition. Two work function values were reported for Be\cite{Haynes_2014_crc, Mechtly_EA_2002}
}
\end{figure}

We examined this observation can by considering one of the energy contributions to the formation of a metal-oxide (ionic ones), the transfer of electrons from the  metal Fermi energy to the oxygen $p$ orbitals (see inset of Fig.\ \ref{fig:Eform_EWF}). The idea behind is that when a metal oxidizes, electrons are transferred from Fermi energy of a metal to oxygen unfilled $p$ orbitals and the energy difference can largely represent the thermodynamic driving force for the oxidation reaction to happen. This could explain why metals with low work functions might be easier to oxidize.

To check how this hypothesis corresponds to the observed trends we constructed a scatter plot of the enthalpies of formation of binary oxides against the work functions of the corresponding metals shown in  Fig.~\ref{fig:Eform_EWF}. We assumed that the energy position of oxygen $p$-orbitals within oxides remain the same for all the oxides. Additionally, some metal can form oxides with different stoichiometries and we include in our analysis only those that have nominal oxidation state of -2 for oxygen (no peroxides for example). This criteria ensure that oxygen captures two electrons only from metals. Formation enthalpies are normalized by the number of oxygen atoms in the chemical formulae.

Fig.~\ref{fig:Eform_EWF} shows a reasonable trend between oxide formation enthalpy per oxygen and work functions of corresponding metals. As expected the trend is qualitative as there is much more to the oxide formation energies than just the electron transfer. The only strong outliers are alkali oxides Na$_{2}$O, K$_{2}$O, and Rb$_{2}$O out of a total of 57 oxides considered. Nonetheless, the trend from Fig.~\ref{fig:Eform_EWF} provides a simple and intuitive way to understand our results and a quick way to estimate whether a given metal will oxidize in contact with an oxide formad by a different metal by comparing the work functions of the two metals. Namely, if the metal in question has a lower work function that the one in an oxide, the formation of the oxide at the interface rather than a stable coexistence with the metal in question is likely. In addition, this chemical intuition is consistent with the known linear correlation between electronegativity and work function for metals\cite{Chen_JCP_1977}.

Lastly, we collected metal work functions from CRC Handbook of Chemistry and Physics\cite{Haynes_2014_crc} and these numbers are known to be sensitive to surface orientations. We used the average values if no values exist for polycrystalline samples. Technetium has no experimentally measured values but has an estimated value at 4.82 eV\cite{Drummond_osti_1999}.
We also note that a very different value from 5.0 eV for Beryllium (3.9 eV) was also reported\cite{Mechtly_EA_2002}.

\section{Conclusion}
%
In summary, we applied the computational (first-principles) thermodynamics to estimate the thermochemical stability of metal/semiconductor interfaces for $\beta$-Ga$_{2}$O$_{3}$, rutile GeO$_{2}$, and wurtzite GaN.  In general, for the two wide-band-gap oxides, we find that only noble metals can form stable contacts under reasonable oxygen partial pressure ($> 10^{-10}$ atm). These metals tend to form Schottky barriers with $n$-type doped $\beta$-Ga$_{2}$O$_{3}$ and rutile GeO$_{2}$ based on the Schottky-Mott rule.  Hence, degenerate doping of active materials, if possible, should be used to lower the interfacial resistance and form effective Ohmic contacts. Alternatively, SnO$_2$ is predicted to stably coexist with both $\beta$-Ga$_{2}$O$_{3}$, rutile GeO$_{2}$ at wide range of temperatures and oxygen partial pressures, which could provide another solution to the Ohmic contact problem for ultra-wide-gap oxides. In comparison, for GaN, consistent with the general knowledge that N$_{2}$ molecules are chemically relatively inert, more metals are predicted to form stable contacts, without involvement of oxygen. The commonly used Ohmic contact metal Ti, is predicted to react with GaN but the products are all conductive nitrides. The same is true for Ta, Mo, and to some extent Zr. Beyond finding stable contacts, generated phase diagrams are also useful for identifying appropriate synthesis condition of ternary compounds.
\vspace{0.5cm}

\section{Acknowledgments}
%
This work was authored by the National Renewable Energy Laboratory (NREL), operated by Alliance for Sustainable Energy, LLC, for the US Department of Energy (DOE) under Contract No. DE-AC36-08GO28308. Funding provided by the Office of Energy Efficiency and Renewable Energy (EERE), Advanced Manufacturing Office (application to contact stability), and by the Laboratory Directed Research and Development (LDRD) Program at NREL (thermochemistry method development). The research was performed using computational resources sponsored by the Department of Energy’s Office of Energy Efficiency and Renewable Energy, located at the National Renewable Energy Laboratory. The views expressed in the article do not necessarily represent the views of the DOE or the US Government.

\bibliography{thermo_contacts}

\end{document}